\RequirePackage{fix-cm}
\documentclass{birkjour_t2}          
\usepackage{graphicx}
\usepackage{amssymb} 

\usepackage{subfigure}
\usepackage{amsmath}
\usepackage{amsfonts}
\usepackage{amsbsy}
\usepackage{times}
\usepackage{mathptmx}
\usepackage[left]{lineno}

\usepackage{color} 

\begin{document}

\title[An analytical comprehensive solution for the superficial waves appearing in gravity-driven flows...]{An analytical comprehensive solution for the superficial waves appearing in gravity-driven flows of liquid films
}


\author{Bruno Pelisson Chimetta}
\address{School of Mechanical Engineering, UNICAMP - University of Campinas\br
Rua Mendeleyev, 200, CEP: 13083-860\br
Campinas, Brazil}
\email{brunopc@fem.unicamp.br}

\author{Erick de Moraes Franklin}
\address{Corresponding author\br
School of Mechanical Engineering, UNICAMP - University of Campinas\br
Rua Mendeleyev, 200, CEP: 13083-860\br
Campinas, Brazil\br
Tel.: +55-19-35213375\br
orcid.org/0000-0003-2754-596X}
\email{franklin@fem.unicamp.br}


\date{January 2, 2020}

\maketitle

\begin{abstract}
This paper is devoted to analytical solutions for the base flow and temporal stability of a liquid film driven by gravity over an inclined plane when the fluid rheology is given by the Carreau-Yasuda model, a general description that applies to different types of fluids. In order to obtain the base state and critical conditions for the onset of instabilities, two sets of asymptotic expansions are proposed, from which it is possible to find four new equations describing the reference flow and the phase speed and growth rate of instabilities. These results lead to an equation for the critical Reynolds number, which dictates the conditions for the onset of the instabilities of a falling film. Different from previous works, this paper presents asymptotic solutions for the growth rate, wavelength and celerity of instabilities obtained without supposing \textit{a priori} the exact fluid rheology, being, therefore, valid for different kinds of fluids. Our findings represent a significant step toward understanding the stability of gravitational flows of non-Newtonian fluids.\\

\noindent \textbf{This is a pre-print of an article published in Zeitschrift f\"{u}r angewandte Mathematik und Physik, 71, 122(2020). The final authenticated version is available online at: https://doi.org/10.1007/s00033-020-01349-x}

\end{abstract}

\section{Introduction}

The study of gravity-driven flows of liquids and suspensions has been the object of interest of many authors through the last decades, not only because of practical engineering applications, such as friction-reducing effects, reactor cooling or coating processes, but also because of different geophysical flows such as mud, glaciers and lava flows. In the case of Newtonian fluids, this class of problems has been extensively studied for almost a century. Among the first studies, Kapitza and Kapitza \cite{kapitsa1948wave,kapitza1949wave} carried out experimental and analytical investigations of Newtonian films flowing in the presence of a vertical wall. Over the last decades, several authors studied falling films by using asymptotic approaches for long and short waves \cite{benjamin1957wave,benney1966long,yih1963stability,lin1946stability}, while others carried out numerical investigations in order to increase accuracy and obtain new results \cite{orszag1971accurate,floryan1987instabilities}. These works established the foundations of the stability analysis of gravity-driven flows of Newtonian films. However, the rheological behavior of a wide class of fluids cannot be properly described by the Newtonian constitutive equation. To overcome this difficulty, it is possible to use a generalized Newtonian fluid, for which the viscosity $\eta$ is a function of the shear rate $\dot{\gamma}$. This class of fluids satisfies the following constitutive equation \cite{morrison2001,macosko1994}:

\begin{equation}\label{eq1}
	\underline{\underline{\tau}} = \eta (\dot{\gamma})\underline{\underline{\dot{\gamma}}}
\end{equation}

\noindent where $\eta(\dot{\gamma})$ is a scalar function and $\dot{\gamma} = |\underline{\underline{\dot{\gamma}}}|$. To describe the viscosity behavior, there are different models according to the considered fluid, some examples being the power-law \cite{ostwald1929}, Bingham \cite{bingham1916} and Cross (Carreau) \cite{cross1965} fluids. For the modeling of a liquid layer in the presence of free surface, the Carreau model is usually considered appropriate because it predicts that the viscosity remains finite while the shear rate approaches zero. However, some authors have preferred a power-law model for the same kind of problem \cite{ng1994,ruyer2012,noble2013}. In this work, we use the Carreau-Yasuda model \cite{yasuda1981}, which is a more general model compared to others, encompassing, for instance, the power-law and Carreau models. The viscosity in the Carreau-Yasuda model is given by Eq. (\ref{eq2}),

\begin{equation}\label{eq2}
	\eta (\dot{\gamma}) = \eta _{\infty} + (\eta _{0} - \eta _{\infty}) [1 + (\dot{\gamma} \lambda) ^{a}] ^{\frac{n - 1}{a}}
\end{equation}

\noindent where, concerning the behavior of viscosity with respect to the applied shear rate, the parameter $a$ controls the shape of the transition region between the zero-shear-rate plateau and the rapidly decreasing (power-law-like) region, parameter $\lambda$ determines the values of the shear rate at which transitions occur from the zero-shear-rate plateau to power-law region and from the power-law region to that where $\eta = \eta _{\infty}$, and exponent $n$ governs the inclination of the rapidly decreasing portion of the $\eta$ curve \cite{morrison2001}. In this model, the viscosity tends to $\eta _{\infty}$ as the shear rate $\dot{\gamma}$ becomes larger, and when the shear rate gets smaller the viscosity tends to $\eta _{0}$. The previous works on now-Newtonian stability using this model fixed the values of some of its parameters \cite{nouar2007,weinstein1990,rousset2007}, limiting the validity of the obtained results. In another way, by considering the parameters variable (mainly $a$), the problem becomes mathematically more difficult, but more general. Different from previous works, this paper presents asymptotic solutions for the growth rate, wavelength and celerity of instabilities obtained without fixing values for $a$, $\lambda$ or $n$, being, therefore, valid for different kinds of fluids.

\begin{figure}[!h]
	\centering
	\includegraphics[scale=0.4]{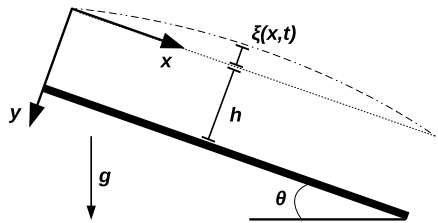}
	\caption{Layout of the falling film over an inclined plane. The main parameters are shown in the figure.}
	\label{fig0}
\end{figure}

In this study, we consider as base state a two-dimensional flow of a liquid film driven by gravity over an inclined plane with slope angle $\theta$, the amplitude of the film undulations being $\hat{\xi}$. Fig. \ref{fig0} presents a layout of the problem and the involved parameters. For Newtonian fluids, it is possible to apply the result of Squire's theorem \cite{squire1933}, even for free-surface \cite{yih1955} and stratified flows \cite{hesla1986}. Gupta and Rai \cite{gupta1968}, making use of the Rivlin-Ericknen model to describe a visco-elastic fluid, showed that under certain circumstances the Squire's Theorem would not be valid for non-Newtonian fluids. However, Nouar et al. \cite{nouar2007}, by considering the Carreau model, found that the two-dimensional instabilities are dominant even for non-Newtonian fluids, so that the Squire's Theorem would be valid. Therefore, we consider in the following that the Squire's Theorem is valid for non-Newtonian fluids and two-dimensional perturbations are dominant. The governing equations are the mass and momentum equations, given in dimensionless form by Eqs (\ref{eq12}) and (\ref{eq13})--(\ref{eq14}), respectively,

\begin{equation}\label{eq12}
	\frac{\partial \overline{u}}{\partial \overline{x}} + \frac{\partial \overline{v}}{\partial \overline{y}} = 0
\end{equation}

\begin{equation}\label{eq13}
	\frac{\partial \overline{u}}{\partial \overline{t}} + \overline{u} \frac{\partial \overline{u}}{\partial \overline{x}} + \overline{v} \frac{\partial \overline{u}}{\partial \overline{y}} = - \frac{\partial \overline{p}}{\partial \overline{x}} + \frac{1}{Re} \bigg(\frac{\partial \overline{\tau} _{xx}}{\partial \overline{x}} + \frac{\partial \overline{\tau} _{xy}}{\partial \overline{y}}\bigg) + \frac{1}{Fr _{x} ^{2}}
\end{equation}

\begin{equation}\label{eq14}
	\frac{\partial \overline{v}}{\partial \overline{t}} + \overline{u} \frac{\partial \overline{v}}{\partial \overline{x}} + \overline{v} \frac{\partial \overline{v}}{\partial \overline{y}} = - \frac{\partial \overline{p}}{\partial \overline{y}} + \frac{1}{Re} \bigg(\frac{\partial \overline{\tau} _{xy}}{\partial \overline{x}} + \frac{\partial \overline{\tau} _{yy}}{\partial \overline{y}}\bigg) + \frac{1}{Fr _{y} ^{2}}
\end{equation}

\noindent  where the dimensionless stress tensor components, viscosity, and shear rate are given by Eqs. (\ref{eq15}), (\ref{eq16}) and (\ref{eq17}), respectively. The other dimensionless quantities are given in Eq. (\ref{eq8}),

\begin{equation}\label{eq15}
	(\overline{\tau} _{xx},\overline{\tau} _{yy},\overline{\tau} _{xy}) = \bigg(2 \overline{\eta} (\overline{\dot{\gamma}}) \frac{\partial \overline{u}}{\partial \overline{x}} \ \ , \ \ 2 \overline{\eta} (\overline{\dot{\gamma}}) \frac{\partial \overline{v}}{\partial \overline{y}} \ \ , \ \ \overline{\eta} (\overline{\dot{\gamma}}) \bigg[\frac{\partial \overline{u}}{\partial \overline{y}} + \frac{\partial \overline{v}}{\partial \overline{x}}\bigg] \bigg)
\end{equation}

\begin{equation}\label{eq16}
	\overline{\eta} (\overline{\dot{\gamma}}) = I + (1 - I) [1 + (L \dot{\gamma}) ^{a}] ^{\frac{n - 1}{a}}
\end{equation}

\begin{equation}\label{eq17}
	\overline{\dot{\gamma}} = \bigg\{ 2 \bigg(\frac{\partial \overline{u}}{\partial \overline{x}} \bigg)^{2} + \bigg[\bigg(\frac{\partial \overline{u}}{\partial \overline{y}} \bigg)^{2} + 2\frac{\partial \overline{u}}{\partial \overline{y}} \frac{\partial \overline{v}}{\partial \overline{x}} + \bigg(\frac{\partial \overline{v}}{\partial \overline{x}} \bigg)^{2} \bigg] + 2 \bigg(\frac{\partial \overline{v}}{\partial \overline{y}} \bigg)^{2} \bigg\}^{\frac{1}{2}}
\end{equation}

\begin{equation}\label{eq8}
(\overline{x},\overline{y},\overline{u},\overline{v},\overline{t},\overline{p},\overline{\eta},I,L) = \bigg(\frac{x}{h_{s}},\frac{y}{h_{s}}, \frac{uh_{s}}{Q},\frac{vh_{s}}{Q}, \frac{t Q}{h_{s}^{2}},\frac{ph_{s}^{2}}{\rho Q^{2}},\frac{\eta}{\eta _{0}}, \frac{\eta _{\infty}}{\eta _{0}}, \frac{\lambda Q}{h_{s}^{2}} \bigg)
\end{equation}

\noindent where $x$ and $y$ are the longitudinal and normal coordinates, $u$ and $v$ are the longitudinal and normal velocity components, $t$ is the time, $p$ is the pressure, $Q$ is the volumetric flow rate by unit of width, $I$ is the ratio between the largest and smallest viscosities, $L$ is a dimensionless relaxation time, and $\rho$ is the fluid density. The characteristic scale is the film thickness of the reference flow, $h_{s}$, given by Eq. (\ref{eq11}) \cite{weinstein1990},

\begin{equation}\label{eq11}
h_{s} = \bigg[\frac{\eta_{0}Q}{\rho g \sin(\theta)} \bigg]^{\frac{1}{3}} 
\end{equation}

\noindent and the pertinent dimensionless groups are the Reynolds and Froude numbers, defined in Eq. (\ref{eq18}).

\begin{equation}\label{eq18}
	(Re, Fr_{x}, Fr_{y}) = \bigg( \frac{\rho Q}{\eta _{0}}, \sqrt{\frac{Q ^{2}}{gh_{s} ^{3}\sin(\theta)}}, \sqrt{\frac{Q ^{2}}{gh_{s} ^{3}\cos(\theta)}}\bigg)
\end{equation}

The base flow is parallel and steady, and the base velocity profile $\overline{u} = \overline{U}$ is a function of $\overline{y}$ only, the normal component of the velocity being equal to zero. The pressure gradient in the $x$-direction is also zero. The boundary conditions are no-slip at the wall ($\overline{y} = \overline{h}$) and zero shear at the free surface ($\overline{y} = 0$). Under these considerations,

\begin{equation}\label{eq22}
	\overline{U}(\overline{y}) = 0 \  \ at \  \ \overline{y} = \overline{h}
\end{equation}

\begin{equation}\label{eq24}
	\bigg\{ I + (1 - I) \bigg[1 + \bigg(L \frac{d \overline{U}}{d \overline{y}} \bigg) ^{a} \bigg] ^{\frac{n - 1}{a}} \bigg \} \frac{d \overline{U}}{d \overline{y}} = - \overline{y} \   \ for \   \ \overline{y} \in [0;\overline{h}]
\end{equation}

\noindent where $\overline{h} = h/h _{s}$. A final relation is obtained by considering the dimensionless flow rate equal to unity. Therefore,

\begin{equation}\label{eq25}
	\int_{0}^{\overline{h}} \overline{U}d \overline{y} = 1
\end{equation}

Eqs. ($\ref{eq22}$), ($\ref{eq24}$) and ($\ref{eq25}$) establish a nonlinear problem for the velocity profile and the film thickness. There is no analytical solution in the general case for this system.

\section{Temporal stability analysis}

In order to study the linear stability of the flow, we make use of small perturbations (dimensionless) for the velocity $(\hat{u},\hat{v})$ and pressure $(\hat{P})$, together with the interface disturbance $(\hat{\xi})$, and neglect all terms $\mathcal{O}(\varepsilon^2)$, where $\mathcal{O}(\varepsilon)$ = $\mathcal{O}(\hat{u}, \hat{v}, \hat{P})$ and the symbol $\mathcal{O}$ refers to the asymptotic order \cite{VanDyke}. Since we consider two-dimensional disturbances, a stream function $\hat{\Psi}$, given by Eq. (\ref{eq55}), can be used.

\begin{equation}\label{eq55}
(\hat{u},\hat{v}) = \bigg(\frac{\partial \hat{\Psi}}{\partial \overline{y}}, - \frac{\partial \hat{\Psi}}{\partial \overline{x}} \bigg)
\end{equation}

\noindent The normal modes for the stream function and interface fluctuation are given by Eqs. ($\ref{eq56}$) and ($\ref{eq57}$),

\begin{equation}\label{eq56}
\hat{\Psi}(\overline{x},\overline{y},\overline{t}) = \tilde{\Psi}(\overline{y}) e ^{i \alpha (\overline{x} - c \overline{t})}
\end{equation}

\begin{equation}\label{eq57}
\hat{\xi}(\overline{x},\overline{t}) = \tilde{\xi} e ^{i \alpha (\overline{x} - c \overline{t})}
\end{equation}

\noindent where $\alpha = kh_{s} \in \mathbb{R}$, $k$ corresponding to the wave number, and $c = \omega h_{s}k^{-1}Q^{-1} \in \mathbb{C}$, $\omega$ being the complex frequency. Inserting Eqs. (\ref{eq56}) and (\ref{eq57}) into Eqs.(\ref{eq12}) to (\ref{eq15}), we find the Orr-Sommerfeld equation for the Carreau-Yasuda model,

\begin{eqnarray*}
(D ^{2} + \alpha ^{2})[D ^{2} \overline{\epsilon} + 2D\overline{\epsilon}D + \overline{\epsilon}(D ^{2} + \alpha ^{2})]\tilde{\Psi} - 4\alpha ^{2}D(\overline{\eta}D\tilde{\Psi}) =
\end{eqnarray*}

\begin{eqnarray}\label{eq58}
= i\alpha Re [(\overline{U}-c)(D ^{2} - \alpha ^{2}) - D ^{2}\overline{U}]\tilde{\Psi}
\end{eqnarray}

\noindent where $D ^{j} = \frac{\partial ^{j}}{\partial \overline{y} ^{j}}$ and $\overline{\epsilon}$ is found by inserting Eq.(\ref{eq55}) into Eq.(\ref{eq17}), leading to

\begin{equation}\label{eq48}
\overline{\epsilon} = I + (1 - I)\bigg[1 + n \bigg(L \frac{\partial \overline{U}}{\partial \overline{y}} \bigg)^{a} \bigg] \bigg[1 + \bigg(L \frac{\partial \overline{U}}{\partial \overline{y}} \bigg)^{a} \bigg]^{\frac{n-a-1}{a}}
\end{equation}

The no-slip condition at the solid boundary is given by Eqs. (\ref{eq59}) and (\ref{eq60}).

\begin{equation}\label{eq59}
D\tilde{\Psi} = 0 \  \ at \  \ \overline{y} = \overline{h}
\end{equation}

\begin{equation}\label{eq60}
\tilde{\Psi} = 0 \  \ at \  \ \overline{y} = \overline{h}
\end{equation}

The boundary conditions at the free surface are the kinematic condition, which represents the impermeability of the interface, and two dynamic conditions, that are related to the continuity of the tangential and normal stresses at the interface, representing the viscous effect and Laplace-Young equation, respectively. Combining these three conditions, we find the free-surface conditions,

\begin{equation}\label{eq64}
[1 + (\overline{U} - c)(D^{2}+\alpha ^{2})]\tilde{\Psi} = 0 \  \ at \  \ \overline{y} = 0
\end{equation}

\begin{eqnarray*}
i\alpha Re[(c - \overline{U})D + D\overline{U}]\tilde{\Psi} - 4\alpha ^{2}\overline{\eta}D \tilde{\Psi} +
\end{eqnarray*}

\begin{eqnarray}\label{eq65}
+ (D^{2}+\alpha ^{2})\bigg[D\overline{\epsilon} + \overline{\epsilon}D + i\alpha \overline{\epsilon} \bigg(\cot(\theta) + \frac{\alpha ^{2}}{We_{m}} \bigg) \bigg]\tilde{\Psi} = 0 \  \ at \  \ \overline{y} = 0
\end{eqnarray}

\noindent where $We _{m} = \eta _{0} Q (h_{s} \gamma )^{-1}$ is a modified Weber number. The system consisting of Eqs.(\ref{eq58}) to (\ref{eq65}) forms a generalized eigenvalue problem, from which we find the celerity $c$ and wave number $\alpha$.

\section{Asymptotic solutions for a Carreau-Yasuda fluid}

\subsection{Base flow and film thickness}

We find next an asymptotic solution for the base flow by considering the limit case for small non-Newtonian behavior when $L$ tends to zero. We write the solutions for the velocity profile and film thickness as exponential expansions given by Eqs. (\ref{eq26}) and (\ref{eq27}), respectively,

\begin{equation}\label{eq26}
	\overline{U}(\overline{y}) = \overline{U}_{0} + \overline{U}_{1}L ^{a} + \mathcal{O}(L ^{m}) \  \ with \  \ m > a
\end{equation}

\begin{equation}\label{eq27}
	\overline{h} = \overline{h}_{0} + \overline{h}_{1}L ^{a} + \mathcal{O}(L ^{m}) \  \ with \  \ m > a
\end{equation}

\noindent and insert them into the system consisting of Eqs. (\ref{eq22}) to (\ref{eq25}). For the zeroth order, $\mathcal{O}(L ^{0})$, we obtain Eqs. (\ref{eq28}) to (\ref{eq30}).

\begin{equation}\label{eq28}
	\overline{U}_{0} = 0 \  \ at \  \ \overline{y} = \overline{h}_{0}
\end{equation}

\begin{equation}\label{eq29}
	\bigg\{ I + (1 - I) \bigg[1 + \bigg(L \frac{d \overline{U}_{0}}{d \overline{y}} \bigg) ^{a} \bigg] ^{\frac{n - 1}{a}} \bigg \} \frac{d \overline{U}_{0}}{d \overline{y}} = - \overline{y}
\end{equation}

\begin{equation}\label{eq30}
	\int_{0}^{\overline{h}_{0}} \overline{U}_{0}d \overline{y} = 1
\end{equation}

In order to solve the above system, it is necessary to neglect any term equal or above $\mathcal{O}(L ^{a})$, and afterward integrate Eq.(\ref{eq29}) by considering Eq.(\ref{eq28}). The result is then inserted into Eq.(\ref{eq30}), and this procedure leads to the solution for $\mathcal{O}(L ^{0})$, given by Eqs. (\ref{eq31}) and (\ref{eq32}).

\begin{equation}\label{eq31}
	\overline{U}_{0}(y) = \frac{(\overline{h}_{0} ^{2} - \overline{y}^{2})}{2}
\end{equation}

\begin{equation}\label{eq32}
	\overline{h}_{0} = \sqrt[3]{3}
\end{equation}

\indent Proceeding next with the terms of $\mathcal{O}(L ^{a})$, we obtain the system consisting of Eqs. (\ref{eq33}) to (\ref{eq35}),

\begin{equation}\label{eq33}
	\overline{U}_{0} + \overline{U}_{1} L^{a} = 0 \  \ at \  \ \overline{y} = \overline{h}_{0} +  \overline{h}_{1} L^{a}
\end{equation}

\begin{equation}\label{eq34}
	\bigg\{ I + (1 - I) \bigg[1 + \bigg(L \frac{d \overline{U}_{0}}{d \overline{y}} + L^{a+1} \frac{d \overline{U}_{1}}{d \overline{y}} \bigg) ^{a} \bigg] ^{\frac{n - 1}{a}} \bigg \} \bigg( \frac{d \overline{U}_{0}}{d \overline{y}} + \frac{d \overline{U}_{1}}{d \overline{y}} L^{a} \bigg) = - \overline{y}
\end{equation}

\begin{equation}\label{eq35}
	\int_{0}^{\overline{h}_{0} + \overline{h}_{1}L^{a}} ( \overline{U}_{0} + \overline{U}_{1} L^{a} d ) \overline{y} = 1
\end{equation}

\indent and, following the same procedure used for the $\mathcal{O}(L ^{0})$ terms, we find

\begin{equation}\label{eq36}
	\overline{U}_{1}(\overline{y}) = \frac{(-1)^{a}(1 - I)(1 - n)(\overline{h}_{0} ^{a+2} - \overline{y} ^{a+2})}{a(a + 2)} + \overline{h}_{0}\overline{h}_{1}
\end{equation}

\begin{equation}\label{eq37}
	\overline{h}_{1} = \frac{(-1)^{a+1}(1 - I)(1 - n) \overline{h}_{0}^{a+1}}{a(a + 3)}
\end{equation}

Combining Eqs. (\ref{eq31}) and (\ref{eq36}) with Eq.(\ref{eq26}), and Eqs. (\ref{eq32}) and (\ref{eq37}) with Eq.(\ref{eq27}), yields the asymptotic solution for the velocity profile and film thickness of the base state for a small non-Newtonian behavior. These equations are a generalization of the solution presented by Weinstein \cite{weinstein1990}, which is a special case for $a=2$ (for $a$ =2, Eqs.(\ref{eq31}), (\ref{eq32}), (\ref{eq36}) and (\ref{eq37}) become exactly the same as Eqs. (A3a), (A3b), (A4a) and (A4b) of Ref. \cite{weinstein1990}).

\subsection{Perturbations and marginal stability}

We present next the zeroth and first order solutions for long-wavelength instabilities, obtained by expanding $\Psi$ and $c$ as power series of the wavenumber $\alpha$, with $\alpha\,\rightarrow\,0$,

\begin{equation}\label{eq66}
\tilde{\Psi} = \tilde{\Psi} _{0} + \alpha \tilde{\Psi} _{1} + \mathcal{O}(\alpha ^{2})
\end{equation}

\begin{equation}\label{eq67}
c = c _{0} + \alpha c _{1} + \mathcal{O}(\alpha ^{2})
\end{equation}

\noindent and inserting Eqs. (\ref{eq66}) and (\ref{eq67}) into Eqs. (\ref{eq58}), (\ref{eq59})-(\ref{eq65}). By considering that $\tilde{\Psi} = 1$ at $\overline{y} = 0$ at the zeroth order, $\tilde{\Psi} = 0$ at $\overline{y} = 0$ at the first order \cite{rousset2007}, and by using Eqs. (\ref{eq31}), (\ref{eq32}), (\ref{eq36}) and (\ref{eq37}), we obtain the zeroth and first order solutions. The zeroth order solution is given by solving Eqs. (\ref{eq68})-(\ref{eq73}).

\begin{equation}\label{eq68}
\overline{\epsilon}D ^{4}\tilde{\Psi}_{0} + 2D \overline{\epsilon}D ^{3}\tilde{\Psi}_{0} + D^{2} \overline{\epsilon}D ^{2}\tilde{\Psi}_{0} = 0
\end{equation}

\begin{equation}\label{eq69}
D\tilde{\Psi}_{0} = 0 \  \ at \  \ \overline{y} = \overline{h}
\end{equation}

\begin{equation}\label{eq70}
\tilde{\Psi}_{0} = 0 \  \ at \  \ \overline{y} = \overline{h}
\end{equation}

\begin{equation}\label{eq71}
D^{3} \tilde{\Psi}_{0} = 0 \  \ at \  \ \overline{y} = 0
\end{equation}

\begin{equation}\label{eq72}
\tilde{\Psi}_{0} = 1 \  \ at \  \ \overline{y} = 0
\end{equation}

\begin{equation}\label{eq73}
\tilde{\Psi}_{0} + (\overline{U} - c_{0})D^{2}\tilde{\Psi}_{0} = 0 \  \ at \  \ \overline{y} = 0
\end{equation}

A Mathematica script was written in the course of this work to solve Eq. (\ref{eq68}) with its boundary conditions (Eqs. (\ref{eq69})-(\ref{eq73})), as well as first order equations (Eqs. (\ref{eq75})-(\ref{eq80}), presented next), and it is available on Mendeley Data \cite{Supplemental2}. The output for the zeroth order is given by Eq. (\ref{eq74}).

\begin{equation}\label{eq74}
	c_{0} = 3^{\frac{2}{3}} + \frac{(-1)^{a} 3^{\frac{a+2}{3}} (a+1) (I-1)(n-1) L^{a}}{a(a+3)}
\end{equation}

The first order solution is obtained by solving Eq. (\ref{eq75}) with its boundary conditions (Eqs. (\ref{eq76})-(\ref{eq80})), and it is given by Eq. (\ref{eq81}). 

\begin{equation}\label{eq75}
\overline{\epsilon}D ^{4}\tilde{\Psi}_{1} + 2D \overline{\epsilon}D ^{3}\tilde{\Psi}_{1} + D^{2} \overline{\epsilon}D ^{2}\tilde{\Psi}_{1} = iRe[(\overline{U} - c_{0})D^{2}\tilde{\Psi}_{0} - \tilde{\Psi}_{0}D^{2}\overline{U}]
\end{equation}

\begin{equation}\label{eq76}
D\tilde{\Psi}_{1} = 0 \  \ at \  \ \overline{y} = \overline{h}
\end{equation}

\begin{equation}\label{eq77}
\tilde{\Psi}_{1} = 0 \  \ at \  \ \overline{y} = \overline{h}
\end{equation}

\begin{equation}\label{eq78}
D^{3} \tilde{\Psi}_{1} + i \{ Re[(c_{0} - \overline{U})D \tilde{\Psi}_{0}+\tilde{\Psi}_{0}D\overline{U}]+\cot(\theta)D^{2} \tilde{\Psi}_{0} \} = 0 \  \ at \  \ \overline{y} = 0
\end{equation}

\begin{equation}\label{eq79}
\tilde{\Psi}_{1} = 0 \  \ at \  \ \overline{y} = 0
\end{equation}

\begin{equation}\label{eq80}
\tilde{\Psi}_{1} + (\overline{U} - c_{0})D^{2}\tilde{\Psi}_{1} - c_{1}D^{2}\tilde{\Psi}_{0}= 0 \  \ at \  \ \overline{y} = 0
\end{equation}

\begin{eqnarray*}
c_{1} = \frac{i [6 Re - 5 \cot(\theta)] }{5} +
\end{eqnarray*}

\begin{eqnarray}\label{eq81}
+\frac{ i (-1)^{a} 3^{\frac{a+3}{3}} (I-1)(n-1)L^{a}\{2 [15 + a (6a+37)] Re - 5 a (a+5) \cot(\theta)\}}{5a(a+3)(a+5)}
\end{eqnarray}

The initial stability of a falling film of a general liquid described by the Carreau-Yasuda model is determined by Eqs. (\ref{eq74}) and (\ref{eq81}), which are generalizations of the solutions proposed by Rousset et al. \cite{rousset2007} for the especial case where $a$ = 2 (for $a$ = 2, Eqs. (\ref{eq74}) and (\ref{eq81}) become exactly the same as Eqs. (42) and (49) of Ref. \cite{rousset2007}). The critical Reynolds number at the onset of instability is given by Eq. (\ref{eq82}), obtained by considering $c_{i} = \alpha c_{1} = 0$.

\begin{equation}\label{eq82}
Re_{cr} = \frac{5 \cot(\theta)}{6} \biggl\{1 + \frac{(-1)^{a+1} 3^{\frac{a}{3}} [15 + a (3a+22)] (I-1)(n-1)L^{a} }{a (a+3) (a+5)} \biggl\}
\end{equation}

For $a$ = 2, Eq.(\ref{eq82}) becomes exactly the same as Eq. (50) of Rousset et al. \cite{rousset2007}.

\section{Results}

The main novelty of our solution is that it is valid for any fluid obeying the Carreau-Yasuda model, without any supposition made \textit{a priori} on the specific kind of fluid. For this reason, we focus the following analysis on the parameter $a$, its continuous variation being impossible to attain by using solutions of previous studies found in the literature, but now achievable with the proposed solution. Because some values of $a$ lead to complex solutions, we considered only their real part in the cases analyzed in this paper. Besides the graphics presented next, other graphics are available as Supplementary Material \cite{Supplemental} and Matlab scripts for plotting the graphics are available on Mendeley Data \cite{Supplemental2}.

Our comprehensive solution was compared with particular solutions found in the literature \cite{weinstein1990,rousset2007}, and the agreement was excellent. Graphics showing direct comparisons between our solution and that of Rousset et al. \cite{rousset2007} for $a = 2$, $n = 0.8$ and $L = 0.5$ (small non-Newtonian behavior) are available as Supplementary Material \cite{Supplemental}. However, different from previous studies, we can now analyze the flow behavior continuously through different types of fluids.

Part of the non-Newtonian fluids described by the Carreau-Yasuda model are in the range 1 $<$ $a$ $<$ 7, some examples being found in Refs. \cite{peralta2017,khechiba2017,japper2009}. We present next the critical conditions for the appearance of film instabilities by varying the parameters $a$ between 1 and 10 and $n$ between 0 and 1, which corresponds to a progressive variation from shear-thinning to Newtonian flows.

\begin{figure}[!htb]
	\centering
	\includegraphics[scale=0.25]{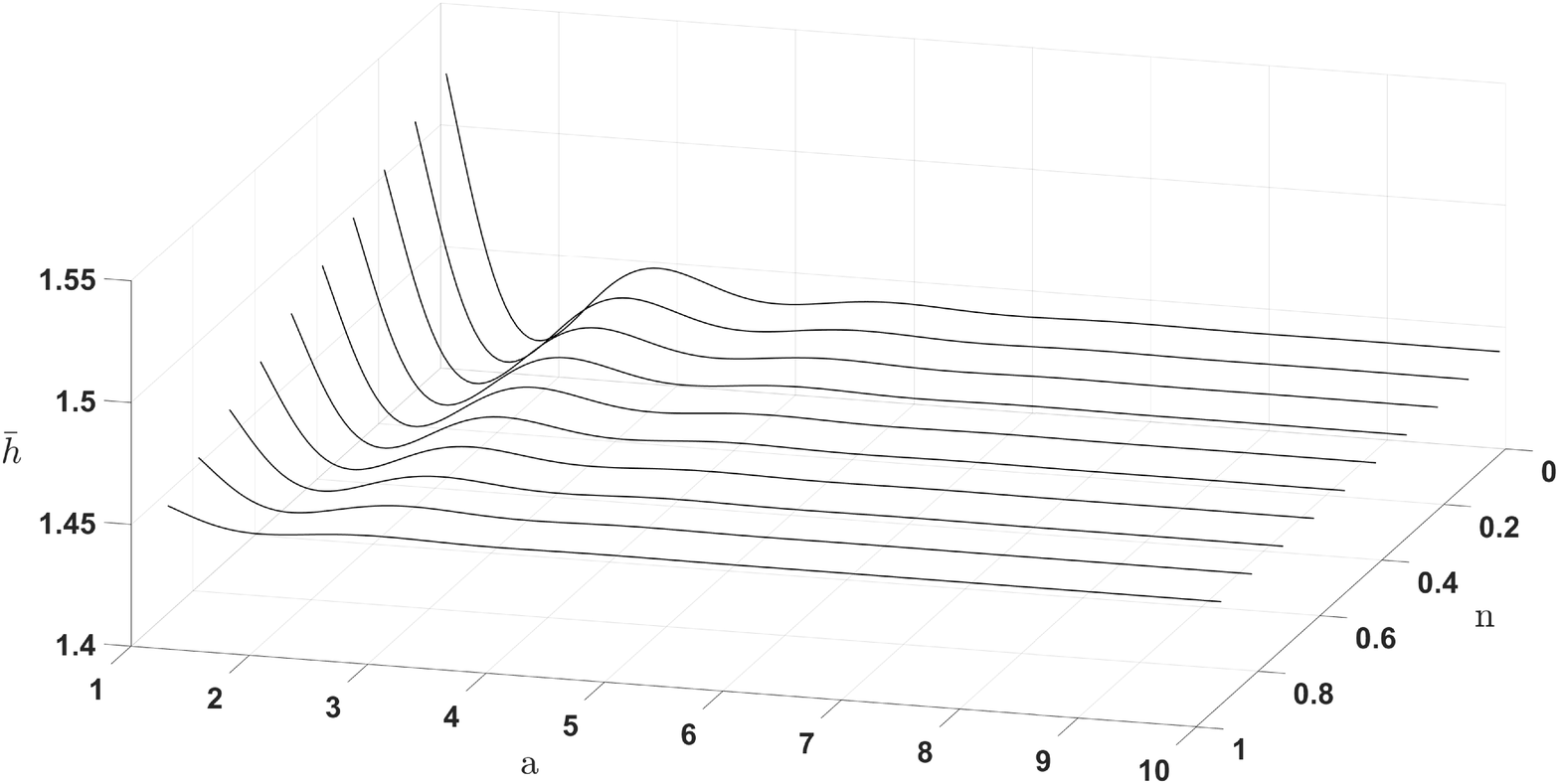}
	\caption{Film thickness $\overline{h}$ as a function of the parameters $n$ and $a$ for a shear-thinning fluid with $L = 0.4$ and $I = 0.5$.}
	\label{fig1}
\end{figure}

Fig. $\ref{fig1}$ presents the film thickness $\overline{h}$ of the base flow as a function of $n$ and $a$, for fixed $I$ and $L$. In this case, our results show that the film thickness increases as the fluid rheology acquires an increasing level of shear-thinning characteristics ($a\,\rightarrow\,1$ and $n\,\rightarrow\,0$). For this specific analysis, the increase in the film thickness with increasing shear-thinning characteristics is due to their higher viscosities for fixed $I$ and $L$ values (as would happen if a high-viscosity shear-thinning fluid and a low-viscosity Newtonian fluid were let to flow on the same inclined plane and their thickness compared). This increase is monotonic with $n$ and non-monotonic with $a$.

\begin{figure}[!htb]
	\centering
	\includegraphics[scale=0.25]{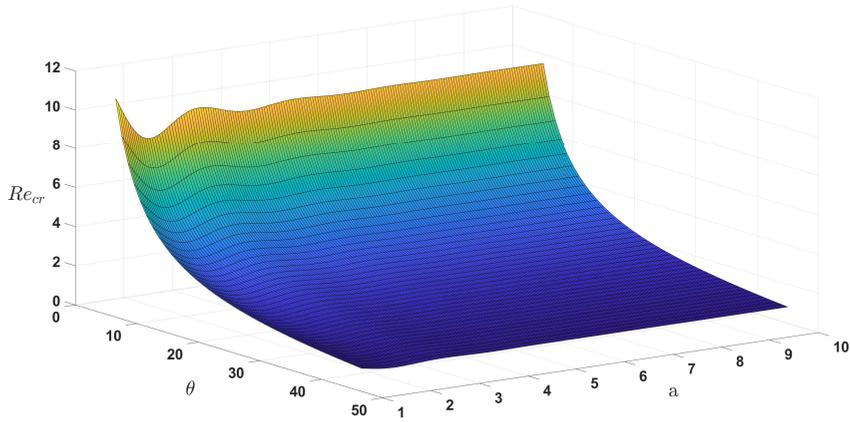}
	\caption{Critical Reynolds number $Re_{cr}$ as a function of the slope angle $\theta$ and $a$ for a shear-thinning fluid with $I = n = 0.5$ and $L = 0.4$.}
	\label{fig2}
\end{figure}

Fig. $\ref{fig2}$ shows the critical Reynolds number $Re_{cr}$ as a function of the slope angle $\theta$ and $a$ for a shear-thinning fluid, investigating further the shear-thinning behavior. From this figure, we observe that the critical Reynolds number decreases with the slope angle, just as happens in the Newtonian case, and that it varies non-monotonically as the fluid rheology tends to stronger shear-thinning characteristics ($a\,\rightarrow\,1$). As the value of $a$ increases further than 7, the shear-thinning effects gradually loose their relevance and the fluid rheology tends to a Newtonian behavior. It is remarkable that fluids with rheology given by $a\,\rightarrow\,1$ are the most stable, while the Cross (or Carreau) fluids, which have $a$ = 2, are the most unstable of shear-thinning fluids: surface waves appear earlier and travel faster for Cross fluids, since in this configuration the liquid film is thinner and surface velocity higher, the opposite happening for $a$ = 1. Those features, observed from continuous variation from Newtonian to shear-thinning fluids, are shown here for the first time. However, the reasons for the non-monotonic behavior of shear-thinning fluids rest to be investigated further.

\section{Conclusion}

This paper presented an analytical solution for the instabilities of gravity-driven flows of liquid films without fixing \textit{a priori} the type of fluid. Different from previous analytical solutions found in the literature, the obtained solution is comprehensive, being valid for fluids as diverse as Newtonian, shear thinning and shear thickening. The base state and perturbations were obtained based on asymptotic expansions, from which the critical conditions at the onset of instability were deduced. Our comprehensive solution was compared with particular solutions found in the literature, and the agreement was excellent, with the difference that now we can analyze the flow behavior continuously through different types of fluids. We investigated the critical conditions for the appearance of film instabilities by varying the parameters $a$ between 1 and 10 and $n$ between 0 and 1, which corresponds to a progressive variation from shear-thinning to Newtoninan flows. Our results have shown that, for fixed $I$ and $L$, the film thickness increases as the fluid rheology acquires an increasing level of shear-thinning characteristics. This increase is monotonic with $n$ and non-monotonic with $a$. Investigating further the shear-thinning behavior, we observed that the critical Reynolds number decreases with the slope angle, just as happens in the Newtonian case, that it varies non-monotonically as the fluid rheology tends to stronger shear-thinning characteristics, and that Carreau fluids are the most unstable of shear-thinning fluids. Our findings represent a significant step toward computing and understanding the behavior of liquid films of non-Newtonian fluids.

\section*{Acknowledgements}
\begin{sloppypar}
This study was financed in part by the Coordena\c c\~ao de Aperfei\c coamento de Pessoal de N\'ivel Superior - Brasil (CAPES) - Finance Code 001. Bruno Pelisson Chimetta is grateful to CAPES, and Erick de Moraes Franklin would like to express his gratitude to FAPESP (Grant No. 2018/14981-7), CNPq (grant no. 400284/2016-2) and FAEPEX/UNICAMP (Grant No. 2112/19) for the financial support they provided.
\end{sloppypar}

\section*{Compliance with ethical standards}
\begin{sloppypar}
\noindent \textbf{Funding} This study was funded by the Coordena\c c\~ao de Aperfei\c coamento de Pessoal de N\'ivel Superior - Brasil (CAPES) - Finance Code 001, Funda\c{c}\~ao de Amparo \`a Pesquisa do Estado de S\~ao Paulo - FAPESP (grant no. 2018/14981-7),  Conselho Nacional de Desenvolvimento Cient\'ifico e Tecnol\'ogico - CNPq (grant no. 400284/2016-2) and Fundo de Apoio ao Ensino, Pesquisa e Extens\~ao da Unicamp - FAEPEX/UNICAMP (grant no. 2112/19).\\

\noindent \textbf{Conflict of interest} The authors declare that they have no conflict of interest.
\end{sloppypar}


\bibliography{references}
\bibliographystyle{spphys}

\end{document}